\documentclass[nobibnotes,showpacs,reprint,prb,amsmath,amssymb,floatfix,superscriptaddress]{revtex4-1}

\usepackage{graphicx}
\usepackage{amsmath} 
\usepackage{xcolor}
\usepackage{upgreek}

\newcommand{\bone}{\mathbf{1}}
\newcommand{\bp}{\mathbf{p}}
\newcommand{\br}{\mathbf{r}}

\newcommand{\bsg}{\boldsymbol{\omega}}
\newcommand{\bxi}{\boldsymbol{\zeta}}

\begin{document}
\title{Tilt induced vortical response and mixed anomaly in inhomogeneous Weyl matter}
\author{Saber Rostamzadeh}
\thanks{saber.rostamzadeh@universite-paris-saclay.fr}	
\affiliation{Laboratoire de Physique des Solides, Universit\'e Paris Saclay, CNRS UMR 8502, 91405 Orsay Cedex, France}
\affiliation{Department of Physics, Istanbul University, Vezneciler, Istanbul, Turkey}
\author{Sevval Tasdemir}
\affiliation{Department of Physics, Istanbul University, Vezneciler, Istanbul, Turkey}
\author{Mustafa Sarisaman}
\affiliation{Department of Physics, Istanbul University, Vezneciler, Istanbul, Turkey}
\author{S. A. Jafari}\affiliation{Department of Physics, Sharif University of Technology, Tehran 11155-9161, Iran}
\author{Mark-Oliver Goerbig}\affiliation{Laboratoire de Physique des Solides, Universit\'e Paris Saclay, CNRS UMR 8502, 91405 Orsay Cedex, France}

\begin{abstract}
We propose a non-dissipative transport effect and vortical response in Weyl semimetals in the presence of spatial inhomogeneities, namely a spatially varying tilt of the Weyl cones. We show that when the spectrum is anisotropic and tilted due to spatial lattice variations, one is confronted with generalized quantum anomalies due to the effective fields stemming from the tilt structure.  In particular, we demonstrate that the position-dependent tilt parameter induces local vorticity, thus generating a \textit{chiral vortical effect} even in the absence of rotation or magnetic fields. As a consequence, it couples to the electric field and thus contributes to the anomalous Hall effect.
\end{abstract}

\maketitle
\section{Introduction}

Dirac materials, among which graphene and Weyl semimetals, provide a natural platform for the prolific exchange of methods and ideas between two distinguished branches of physics, namely high-energy physics, and low-energy condensed-matter physics.\cite{wehling2014dirac}  A concept that is relevant for both branches is definitely that of \textit{gauge theories} that describe the laws of physics which are invariant under local transformations\cite{peskin2018introduction,srednicki2007quantum}. This provides a unifying language to treat low-energy condensed-matter systems where the Dirac fermions are subject to the various perturbations stemming from the background lattice potentials in terms of effective gauge fields\cite{kleinert1989gauge,edelen2012gauge}.

These lattice degrees of freedom are coupled to the fermionic degrees of freedom in Dirac systems and thus alter the dynamics of the fermions. For instance, elastic deformations of the honeycomb lattice of graphene through strain\cite{vozmediano2010gauge,choi2010effects,si2016strain} yield fields that are reminiscent of magnetic fields and therefore create effective Landau levels \cite{levy2010strain,meng2013strain,low2010strain}. Furthermore, they have been shown to induce an energy gap and a transition to the topologically insulating regime\cite{guinea2010energy,choi2010controlling}. Lattice dynamics also affect the electronic properties of the three-dimensional (3D) counterpart of graphene, the so-called Weyl semimetals\cite{cortijo2015elastic,aidelsburger2018artificial}, which host various anomalous phenomena in the absence of external fields\cite{cortijo2016strain,pikulin2016chiral,PhysRevLett.122.056601,heidari2020chiral,liu2013chiral}. Applying an external mechanical uniaxial strain to the crystal perturbs the dispersion relation of the Weyl semimetals making it anisotropic. Moreover, such strain can tilt the dispersion relation, which thus has the form of an anisotropic (Dirac or Weyl cone) with an eccentricity, which, upon the increase, can trigger a transition to a type-II Weyl semimetal\cite{PhysRevB.78.045415,soluyanov2015type}. Such tilted anisotropic conical dispersions of Weyl fermions are a specificity of condensed-matter systems as compared to high-energy physics, where the free (elementary) particles are dynamically constrained by the symmetries of space-time. Indeed, the isotropy of space precludes anisotropic cones or tilts that require singling out a particular direction. On the contrary, in condensed-matter physics, these Weyl fermions emerge in lattice models, and the lattice environment naturally breaks the continuous rotation symmetry. The eccentricity of the Dirac cones causes the Fermi velocity to be asymmetric and the conductivity to be highly anisotropic and direction-dependent\cite{rostamzadeh2019large,trescher2015quantum}. Finally, its orientation with the magnetic field distinguishes the electric from the magnetic regimes\cite{tchoumakov2016magnetic} and is the source of yet another intrinsic mechanism generating the anomalous Hall effect.\cite{zyuzin2016intrinsic}

Beyond the spectrum, the tilt of the Dirac cone affects the transport properties of the system, namely in the form of the \textit{chiral anomaly}, which is ubiquitous in the chiral matter and denotes the imbalance in the electronic density of Weyl nodes with different chirality and the resultant transport effect associated with it.\citep{PhysRevB.96.121116,PhysRevB.95.115410} The chiral anomaly shows, however, a slightly variant behavior in Weyl semimetals with a tilted spectrum. {The nature of the chiral modes, as the source of the anomaly, drastically depends on the angle between the tilt and the magnetic field \cite{udagawa2016field}. Moreover, it affects the dependence of the longitudinal magnetoconductivity and its signature on the tilt\cite{sharma2017chiral}, thus indicating the importance of the tilt on the anomaly-induced transport in tilted Weyl semimetals.}

Recently, Weyl semimetals with inhomogeneous tilt parameter drew much attention as it promises a suitable platform to study emergent gravity and black-hole physics in solid-state settings.\cite{volovik2016black,*nissinen2017type,*volovik2018exotic,faraei2019perpendicular,PhysRevB.100.075113,PhysRevB.99.235150,cortijo2016emergent} In this case, the parameter that controls the tilt of the Dirac cones becomes position-dependent and thus provides a suitable knob to probe the more general anomaly-related phenomena for which the Weyl fermions are the culprit. This involves the detection of the gravitational anomaly\citep{gooth2017experimental,nissinen2020emergent} as well as mixed anomaly\citep{PhysRevLett.122.056601,stone2018mixed} in anisotropic Weyl semimetals. 

In the present paper, we propose yet another exotic electronic behavior of Weyl semimetals subjected to a spatially dependent tilt parameter. Due to its position dependence, we show that the rotational of the tilt parameter induces a vorticity field that acts as a Coriolis force on the Weyl quasiparticles and yields additional terms to the chiral anomaly that we call \textit{vortical anomaly}{ for lack of a better name. Indeed, it should be rather called contribution to the chiral anomaly due to the vorticity of the tilt parameter}. In order to test its physical relevance, we study the transport signatures of this vortical anomaly. Indeed, it generates a chiral vortical effect as an effective topological response due to a tilt profile that shares similarities with global rotations of the system\cite{dayi2017semiclassical,PhysRevLett.109.162001,bacsar2014triangle,kharzeev2016chiral}. In this sense, the generated vorticity field is similar to a magnetic field and produces similar anomalous effects to those associated with an external magnetic field. However, we show that the nature of the effects is different owing to the difference in the coupling of the Weyl fermions to the vorticity and magnetic fields.\cite{PhysRevLett.103.191601,PhysRevLett.106.062301} Moreover, as another transport signature, we find that the tilt and its spatial profile can generate an anomalous transverse effect and propose a setup to detect it. {While our analysis is generally valid for both type-I and type-II Weyl semimetals, it is relevant mainly in the former case. Indeed, the delicate transport features we investigate here may be swamped by the overwhelming response of the large density of states at the Fermi level in type-II Weyl semimetals.}

The paper is organized as follows. In Sec. \ref{sec. 2}, we present the basic model and relate the tilt velocity of a generic time-reversal symmetric Weyl Hamiltonian to the underlying geometry of the parameter space. This allows us to derive the effective Lagrangian, and the resulting equations of motion are presented in Sec. \ref{sec. 3}. Section \ref{sec. 4} then deals with the mixed anomaly due to the position-dependent tilt, and its transport signatures in the form of the chiral vortical effect are discussed in Sec. \ref{sec. 5}. There, we discuss how one might experimentally use circularly polarized light to pump a specific valley with a well-defined chirality, and we conclude the paper in Sec. \ref{sec. 6}.

\section{Model}\label{sec. 2}

\begin{figure}[h]
	\centering
	\includegraphics[width=.8\linewidth]{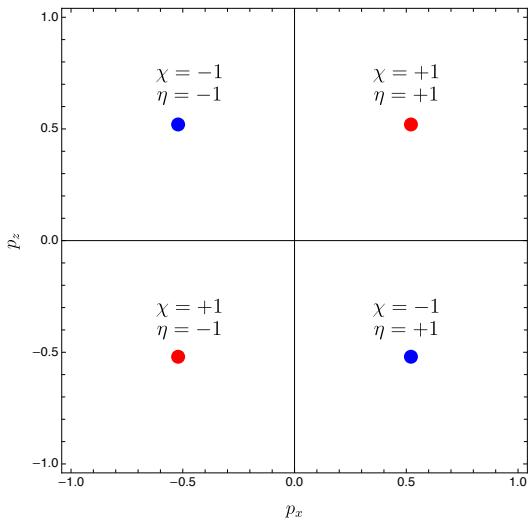}
	\caption{(Color online) Four band contact points in time-reversal symmetric Weyl Hamiltonian. The red (blue) points are TR conjugate of each other and carry equal topological charge $\chi=+1$ ($\chi=-1$).} 
	\label{fig:4nodes}
\end{figure}

To determine the electronic transport properties of Weyl fermions that are generated due to the spatial profile of the tilt term, we consider a time-reversal (TR) symmetric model of the
 Weyl semimetal. 
{Note that when TR symmetry is present, the two Weyl nodes at opposite positions $\bp_W$ and $-\bp_W$ in the first Brillouin zone must have the same topological charges (Chern number). Therefore another pair of Weyl nodes (situated at $\bp_W^\prime$ and $-\bp_W^\prime$) with opposite chirality is required to make the total topological charge of the system vanish. Then the overall effective low-energy Hamiltonian describing two pairs of TR symmetric Weyl nodes with different topological charges is given by (see Appendix A)
\begin{align}\label{Ham}
H_{\chi,\eta}=\eta\:v_x p_x\sigma_x+v_y p_x\sigma_y+\eta\chi\:v_z p_z\sigma_z+\eta\:\mathbf{v}_t\cdot\mathbf{p},
\end{align}
where $p_j$ are the components of the lattice momentum in the continuum limit (with $j=x,y.z$) {and $\boldsymbol{\sigma}=(\sigma_x,\:\sigma_y,\:\sigma_z)$ is the vector of the Pauli matrices in the band space.} {Such continuum Hamiltonian, for instance, can be realized by introducing space-dependent orbital manipulations in 2D systems\citep{yekta2021tune}.} 

Within this model, we consider that the Weyl cones can be generically anisotropic with the velocity parameters $v_j$. However, in the following discussion, we consider an isotropic version $v_j=v_F$, which may always be obtained by rescaling the lattice momenta in an appropriate manner. Furthermore, we have used the velocity parameters $\mathbf{v}_t=v_F\:\bxi(\br)$ in the last term. They describe the tilt of the cones, and we define the tilt parameter and its components $\bxi(\mathbf{r})=(\zeta_x(\br),\zeta_y(\br),\zeta_z(\br))$. {The position dependence of the tilt can generically be obtained from a locally varying strain field in the crystal. This strain field affects the hopping parameters of the tight-binding model underlying the effective low-energy Hamiltonian (\ref{Ham}) and thus the velocity parameters\citep{PhysRevB.78.045415}. The resulting tilt parameter can thus be written in terms of the strain fields as $\zeta_i(\mathbf{u}(\br))=\zeta_i+u_{ij}(\br)\zeta_j$, where the strain tensor $u_{ij}=(\partial_iu_j+\partial_ju_i)/2$ is the symmetrized derivative of the displacement field $\mathbf{u}$\citep{PhysRevLett.108.227205,*PhysRevB.97.201404}. Notice that, at first sight, the position dependence of the tilt parameter renders the Hamiltonian (\ref{Ham}) non-Hermitian, and one would in principle need to replace its last term by the symmetrized version $[\mathbf{v}_t(\mathbf{r})\cdot \mathbf{p}+\mathbf{p}\cdot\mathbf{v}_t(\mathbf{r})]/2$. This amounts to adding the commutator $[p_\mu,v_{t,\mu}(\mathbf{r}]/2=-i\hbar v_F\nabla\cdot\bxi(\br)/2$ to the Hamiltonian. Elastic modification of the tilt profile, moreover, allows for a gauge choice $\nabla\cdot\boldsymbol{\zeta}=0$. As we discuss in more detail Sec. \ref{sec. 4}, this is the case for volume-preserving deformations, in which case the Hamiltonian (\ref{Ham}) is therefore Hermitian without explicit symmetrization of the last term.}

Moreover, TR symmetry requires a tilt parameter in opposite directions for the two Weyl cones situated at opposite momenta so that $\bxi_\eta=\eta\:\bxi$, where the two copies of a TR related pair of Weyl nodes are characterized by the index $\eta=\pm$. The modulus of the tilt parameter recognizes two limits as $\zeta<1$  (mild) and $\zeta>1$  (over tilted) which separates the type-I from type-II Weyl semimetal, respectively, as it has been pointed out first in 2D organic crystals \cite{PhysRevB.78.045415} and then in 3D Weyl semimetals \cite{soluyanov2015type}. The four Weyl nodes are generically situated at four points in the $p_y=0$ plane, as shown in Fig. \ref{fig:4nodes}. Notice that the quartet of Weyl nodes may arise in further copies if certain point symmetries of the lattice are taken into account.}

The tilt term breaks the Lorentz symmetry of the Hamiltonian systems hosting relativistic massless Dirac/Weyl fermions\cite{rostamzadeh2019large}. Most saliently, the tilt term is closely related to the underlying geometry of the parameter space (the momentum space) where the Weyl fermions move. To make this connection, observe that the last term in the Hamiltonian (\ref{Ham}) (considering a single cone) can be understood as a time-like component of a more general 4-velocity tensor such that the Hamiltonian can, in general, be cast into the form $H=v_a^\mu \sigma^a p_\mu$ with the identifications $v_a^\mu=(v_0^\mu,v_1^\mu,v_2^\mu,v_3^\mu)$, wherein this combined picture the latin index pertains to the spin degree of freedom whereas the greek indices show the space-time coordinates. Note that the non-zero elements are only $v_0^\mu=\eta v_F\:\zeta^\mu$ and $\eta v_1^1=v_2^2=\eta v_3^3=v_F$. With this choice, the anisotropy stems only from the tilt parameter $\zeta^\mu=(\bone,\bxi)$, which breaks both rotation and Lorentz symmetries. This also allows one to associate the tilt feature with an emergent underlying metric that affects the dynamics of the Weyl fermions.\cite{Grushin2018} 

In order to further analyze the relationship between the tilt parameter and Lorentz symmetry, it is preferable to switch from the Hamiltonian to the Lagrangian formalism. The Lagrangian density of a tilted Weyl system is given (in a system of units with $\hbar=1$) by \citep{RevModPhys.90.015001}
\begin{equation}\label{lagr}
\mathcal{L}=v_F\;{\psi}^\dagger\left(i\sigma^\mu\partial_\mu+i\bxi(\mathbf{r})\cdot\nabla\right)\psi, 
\end{equation}
where $\psi$ is the Weyl spinor with a definite chirality 
and we have adopted the relativistic notations for the space-time $x^\mu=(v_F t,\mathbf{r})$ and the energy-momentum four-vector $p^\mu=(\frac{1}{v_F}\varepsilon,\mathbf{p})$. The Pauli four-matrices are defined as $\sigma^a=(\bone,\boldsymbol{\sigma})$ and, accordingly, the four-derivative is given by $\partial_\mu=(\frac{1}{v_F}\partial_t,\nabla)$. In the absence of the second (tilt) term, the Lagrangian density 
(\ref{lagr}) is evidently Lorentz covariant, while the tilt term is only given in terms of a three-vector of the spatial components.  

Next, to bring the Lagrangian density 
(\ref{lagr}) into a canonical form, note that the effect of the metric on the Weyl fermions corresponds to an effective metric of a gravitational field\citep{volovik2020vielbein,nissinen2017type,VOLOVIK2022168998} and these effective fields are described by the vielbein (tetrad) $e_i^\mu$ such that
the effective metric reads $g_{\mu\nu}=e_\mu^ae_\nu^b\eta_{ab}$, where $\eta_{ab}$ is the Minkowski metric. This helps us to recast the dispersion into the form $g^{\mu\nu}p_\mu p_\nu=(\varepsilon-v_F\:\bxi\cdot\bp)^2-v_F^2|\bp|^2=0$ and to write the invariant displacement element in the Painlev\'e-Gullstrand coordinate as
\begin{equation}\label{spat_dist}
g_{\mu\nu}dx^\mu dx^\nu=-dt^2+(v_F\:d\mathbf{r}-\bxi dt)^2,
\end{equation}
 where ($\zeta=|\bxi|$) the new metric,
\begin{align}\label{mett_ric}
g_{\mu\nu}=
\left(\begin{array}{l|lll}
1-\zeta^2&-\zeta_x&-\zeta_y&-\zeta_z\\
\hline
-\zeta_x&1&0&0\\
-\zeta_y&0&1&0\\
-\zeta_z&0&0&1
\end{array}\right),
\end{align}
yields the Lagrangian of Dirac equation in the curved space-time
\begin{align}\label{Lagr_curved}
\mathcal{L}=v_F\:{\psi}^\dagger\:(i{\sigma}^a e_a^\mu{D}_\mu)\:\psi . 
\end{align}
This analogy, naturally, bids the adoption of geometrical approaches to study the dynamics of Weyl fermions when spatial perturbations are present in the system. 
This indicates that introducing strain distortion on the system induces a metric tensor and frame fields that intermix the space and time components. The tensor fields $e_a^\mu$ are the Lorentz frame fields (vierbein)\cite{zubkov2018black,nissinen2018tetrads} that redefines the metric. They are related to the elements of the anisotropic 4-velocity matrix, through $v_a^\mu=v_F\:e_a^\mu$, and in a space-time basis they read as
\begin{equation}
e_a^\mu(\mathbf{r})=
\left(\begin{array}{c|c}
\;1\;&-\bxi\\
\hline
\;0\;&1
\end{array}\right).
\end{equation}
Therefore, the tensor fields and thus the metric can be entirely determined by the materials' band structure. The space-time displacement (\ref{spat_dist}) and the underlying effective metric $g_{\mu\nu}$ (\ref{mett_ric}) are reminiscent of the metric near a gravitational source studied in the theories of gravity\cite{volovik2017lifshitz,guan2017artificial,PhysRevResearch.2.043285,PhysRevB.100.075113}, and they have also been noticed in the dynamics of driven Floquet model in Dirac systems. \cite{PhysRevResearch.2.023085}

The covariant derivative in Eq. (\ref{Lagr_curved}) can be decomposed as
\begin{align}
D_\mu=\partial_\mu+\Gamma_\mu-ieA_\mu,
\end{align}
which consists of two main contributions. The first one is the usual electromagnetic vector potential $A_\mu=(\phi,\mathbf{A})$. More important for our study is the spin connection\cite{ryder1996quantum,kleinert1989gauge} $\Gamma_\mu$, which may be viewed as a gauge field generated by the local Lorentz transformation and that naturally arises due to the spatial perturbations present in the Hamiltonian (the tilt parameter) and the anisotropy of the dispersion\cite{PhysRevResearch.2.023410}.

\section{Semiclassical dynamics}\label{sec. 3}

The electronic transport of the carriers, in general, is interpreted as the motion of the particles under the influence of external perturbations, which is described by the equations of motion of the system. To obtain the classical limit of the equations of motion for the Weyl fermions, we consider the transition amplitude between initial $|\bp_i\rangle$ and final state $|\bp_f,t\rangle$. Summing over all degrees of freedom, \textit{i. e.} momentum and spin, give the scattering rates required to model transport. For the Hamiltonian (\ref{Ham}), we employ a path integral formulation and define the propagator as a sum over all the phase-space trajectories as
\begin{equation}
I=\int D[\br]D[\bp]\;
\exp\left(i\int_{t_i}^{t_f}(\bp\cdot\dot{\mathbf{r}}-H)dt\right),
\end{equation}
where the path ordering is implicitly considered. It is useful to reduce the matrix structure of the given action by diagonalizing and projecting onto a single energy band. The diagonalization of the Hamiltonian 
in the chiral basis then yields
\begin{equation}
U_{\chi,\eta}^\dagger\;H\;U_{\chi,\eta}
=(|\mathbf{v}\cdot\mathbf{p}|\sigma_z+\eta\: \bxi\cdot \mathbf{p})=\varepsilon({\mathbf{p}}),
\end{equation}
where $U_{\chi,\eta}(\mathbf{p})$ is a momentum-dependent SU$(2)$ rotation in the pseudo-spin space spanned by the Pauli matrices. Note that during the diagonalization the matrix-dependent part of the kernel is path-ordered,
\[
: e^{-i\delta t \;\mathbf{v}\cdot\mathbf{p}_{n}\cdot\boldsymbol{\sigma}}\;e^{-i\delta t \;\mathbf{v}\cdot\mathbf{p}_{n-1}\cdot\boldsymbol{\sigma}}:
\]
and transforms as
\begin{align}\label{unit_prod}
&U_{\chi,\eta}^\dagger(\mathbf{p}_{n})\left(e^{-i\delta t \;\mathbf{v}\cdot\mathbf{p}_{n}\cdot\boldsymbol{\sigma}}\;e^{-i\delta t \;\mathbf{v}\cdot\mathbf{p}_{n-1}\cdot\boldsymbol{\sigma}}\right)U_{\chi,\eta}(\mathbf{p}_{n-1})=\nonumber\\
&e^{-i\delta t |\mathbf{v}\cdot{\mathbf{p}_{n}}|\sigma_z}[U_{\chi,\eta}^\dagger(\mathbf{p}_{n}) U_{\chi,\eta}(\mathbf{p}_{n-1})]\;e^{-i\delta t |\mathbf{v}\cdot{\mathbf{p}_{n-1}}|\sigma_z},
\end{align}
upon diagonalization of two consecutive terms. 
By power expansion of the unitary transformation around the neighbor point $\mathbf{p}_{n}$ one notes that $U(\mathbf{p}_{n-1})=\exp{(\delta\mathbf{p}\cdot\nabla_\mathbf{p})} U(\mathbf{p}_{n})$, where the infinitesimal interval $\delta\mathbf{p}={\mathbf{p}_{n}}-{\mathbf{p}_{n-1}}$ is related to the time interval via $\delta\mathbf{p}=\dot{\mathbf{p}}\;\delta t$. The middle product in the second line in Eq. (\ref{unit_prod}) thus yields
\begin{equation}
U^\dagger(\mathbf{p}_{n}) U(\mathbf{p}_{n-1})=U^\dagger(\mathbf{p}_{n})e^{\delta\mathbf{p}\cdot\nabla_\mathbf{p}}
U(\mathbf{p}_{n})=e^{-i\delta\mathbf{p}\cdot\mathcal{A}},
\end{equation}
where $\mathcal{A}_{\chi,\eta}(\mathbf{p})=iU_{\chi,\eta}^\dagger(\mathbf{p})\nabla_{\mathbf{p}} U_{\chi,\eta}(\mathbf{p})$ is nothing other than the Berry connection. This gauge term, which naturally arises during the course of the diagonalization, consequently suggests viewing the two opposite chiral Weyl fermions as the monopole and antimonopole of an Abelian theory with strength $\mathfrak{b}=\nabla_\bp\times\mathcal{A}=i\nabla_\bp U^\dagger\times\nabla_\bp U$ and generating Berry field $\mathfrak{b}_z$ (Appendix.A).\cite{PhysRevLett.109.162001}
Therefore, the resulting diagonalized Lagrangian in the presence of electromagnetic potentials reads as\cite{PhysRevB.53.7010,PhysRevLett.75.1348,PhysRevLett.109.162001} 
{
\begin{equation}\label{lag}
\mathcal{L}=(\mathbf{p}+\mathbf{A})\cdot\dot{\mathbf{r}}-(|\mathbf{v}\cdot\mathbf{p}|\sigma_z+\eta\:v_F\bxi\cdot \mathbf{p})-\mathcal{A}\cdot\dot{\mathbf{p}}-\phi.
\end{equation}}
By variation of the Lagrangian, and noting $\dot{\mathbf{A}}(\mathbf{r},t)=\partial\mathbf{A}/\partial t+(\dot{\mathbf{r}}\cdot\nabla_\mathbf{r})\mathbf{A}$ and $\dot{\mathcal{A}}(\mathbf{p})=(\dot{\mathbf{p}}\cdot\nabla_\mathbf{p})\mathcal{A}$, one obtains the equations of motion 
{
\begin{subequations}\label{anom}
\begin{align}\label{vel}
\dot{\mathbf{r}}&=\mathbf{v}_\bp+\eta\:\mathbf{v}_t+\chi\:\dot{\mathbf{p}}\times\mathfrak{b},\\
\label{fors}
\dot{\mathbf{p}}&=\left(-e\:\mathbf{E}+\eta\:[\bp\cdot\nabla]\bxi\right)-e\:\dot{\mathbf{r}}\times\mathbf{B}+\eta\:\bp\times\bsg.
\end{align}
\end{subequations}}
First, note that by turning off the tilt effect ($\bxi=0$), we restore the standard equations of motion\cite{PhysRevLett.109.162001}.
Here the group velocity is given by the sum of the conventional (isotropic) part $\mathbf{v}_\bp=v_F \bp/|\bp|$, the tilt velocity $\mathbf{v}_t$ and the anomalous velocity. In the second equation for $\dot{\bp}$, one notices a generalized electric field term in the parentheses: in addition to the real electric field, there is a second term that stems again from the spatial variation of the tilt term. Here, more precisely, the tilt-velocity gradient $\partial_{r_i}\zeta_j$ is the Jacobian matrix for the tilt parameter\cite{spivak2016magnetotransport} and the curl of the spatially dependent tilt term $\bsg=\nabla_\br\times\bxi$ plays the role of the vorticity.

The tilt-induced vorticity [the last term of Eq. (\ref{fors})] and its consequences are the main results of this paper, so we shall investigate them in more depth. From dimensional analysis, the last term in Eq. (\ref{fors}), indeed, turns out to be the Coriolis force $(\varepsilon/v_F^2) \dot{\br}\times\bsg_\eta$ that is exerted on the particle due to the inhomogeneity of the medium encoded in the tilt term. {In spite of its similarity with the second last term, which bears the chiral anomaly and where one substitutes the magnetic field $\mathbf{B}$ by the vorticity $\bsg$, the latter couples to the momentum rather than the velocity. While the modulus of the velocity remains energy-independent, this is not the case for the momentum, which vanishes at the crossing point. This explains the energy dependence, and consequently its dependence on the chemical potential of the vortical term.}

In general, a particle can feel the Coriolis force, at least in Newtonian limit, around a rotating gravitational source. Its involvement in the dynamics of the Weyl particles in a tilted cone, thus, draws  a strong analogy between the (pseudo)gravitational force and the vorticity $\bsg$, which originated from the tilt profile. If we furthermore interpret the vorticity as a pseudomagnetic field, then the effect is reminiscent of the \textit{gravitomagnetic} effect\cite{ryder1996quantum}.

The identification of the vorticity with a pseudomagnetic field, as a consequence, naturally leads to expect that it can generate a chiral response and thus a chiral vortical effect, which in previous studies is understood as a response of a chiral system to a global rotation.\cite{PhysRevLett.109.162001}

We compute the anomaly-related effect due to the vorticity using the semiclassical Boltzmann equations with Eqs. (\ref{anom}) as ingredients. Using the kinetic equation for the distribution function $f(\mathbf{r},\mathbf{p})$, the transport quantities such as the current density can be computed for a quasiparticle of a definite chirality $\chi$ and TR index $\eta$ by including the weighted phase-space volume in the integration measure as
\begin{align}\label{curr}
\mathbf{J}=-e\int\frac{d\bp}{(2\pi)^3}\:\sqrt{G}\:\dot{\br}\:f(\mathbf{r},\mathbf{p}),
\end{align}
where $\sqrt{G}=1+e\:\chi\:\mathbf{B}\cdot\mathfrak{b}$ is the renormalized volume of the phase-space due to the Berry curvature.

\section{conventional and mixed anomaly}
\label{sec. 4}
The semiclassical Boltzmann equation for the distribution function, which makes use of the equations of motion (\ref{anom}), is given by
\begin{equation}\label{boltz}
\partial_t f+\dot{\mathbf{r}}\cdot\nabla_\br f+\dot{\mathbf{p}}\cdot\nabla_\bp f=-\frac{\delta f}{\tau}.
\end{equation}
By linearization in terms of the electric field $\mathbf{E}$, we find that the small deviation from the equilibrium is due to the elastic disorder scatterings and the electric field, $\delta f=f-f_{\text{eq}}\sim \tau \:\mathcal{O}\cdot\mathbf{v}\:\partial f/\partial\varepsilon$ where $\mathcal{O}(\mathbf{E})$ is a first-order function of the electric field. We assume that the scatterings around a valley relax much faster than the intervalley scattering, $\tau_{\text{inter}}\gg \tau$, thus, the nonequilibrium part of the distribution function is mainly dominated by the intravalley scatterings. By reintroducing the Berry curvature and the dressed velocity through the momentum derivative $\nabla_\bp\rightarrow \dot{\br}\frac{\partial}{\partial\varepsilon}$, then the Boltzmann equation can be integrated to yield the continuity equation for the Weyl fermions around a single Weyl cone ($\tau_\text{inter}\rightarrow\infty$) with the characteristic indices $\chi$ and $\eta$ and the chemical potential $\upmu$, i.e.,

\begin{align}\label{conti}
\partial_t n_{\chi,\eta}+ \nabla_\br\cdot\mathbf{J}_{\chi,\eta}&= \frac{\chi\:e^2}{6\pi^2}(\mathbf{E}\cdot\mathbf{B})+\frac{\chi\eta\:e\:\upmu}{12\pi^2v_F^2}(\mathbf{E}\cdot\mathbf{\bsg})\nonumber\\
&\qquad+\frac{n_{\chi,\eta}}{\tau}.
\end{align}
As it is seen on the right-hand side of Eq.(\ref{conti}) there are two contributions to the quantum anomalies that we will explore in-depth in the following. {As already noticed in the semiclassical equations of motion, the rotational of the tilt parameter plays an analogous role as the magnetic field. However, we insist on the fact that this additional term is not directly generated by an external (magnetic) field as in the usual chiral anomaly. It is rather an indirect effect due to the distortion of the underlying lattice that acts then, via the tight-binding parameters, on the fermionic degrees of freedom. Albeit indirect, the tilt should be seen as an intrinsic property of the fermions, a situation regularly encountered in condensed-matter systems. However, the similarity in the dynamical consequences of this term and that proportional to the usual magnetic field is the reason why we will hence call the resulting effect vortical anomaly.}

\begin{figure}
	\centering
	\includegraphics[width=.9\linewidth]{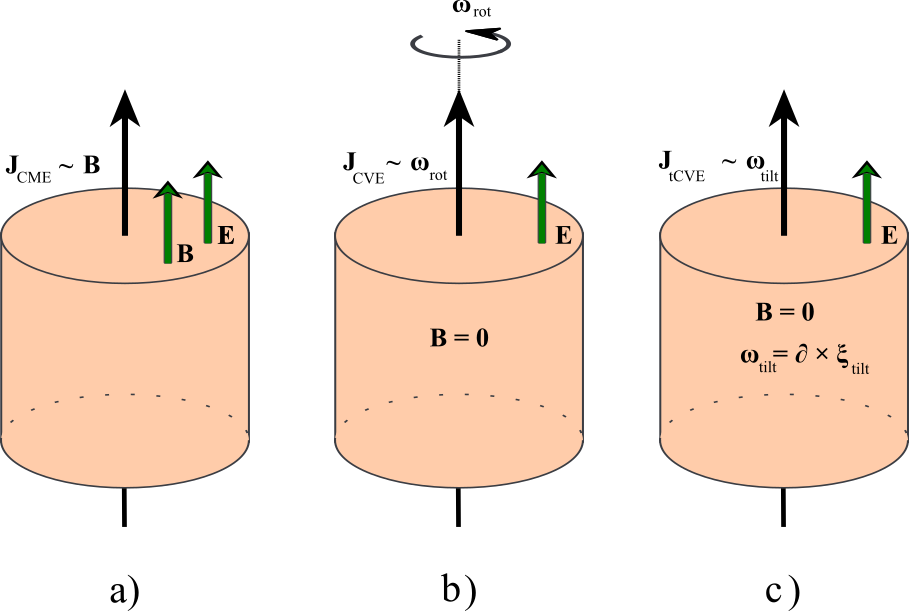}
	\caption{(Color online) Schematic illustration outlining the compendium of anomaly-induced effects: the chiral magnetic effect (CME), the chiral vortical effect (CVE), and the tilt-induced chiral vortical effect (tCVE). (a) Illustrates the standard CME in Weyl matter in the presence of the parallel electric $\mathbf{E}$ and magnetic $\mathbf{B}$ fields where a net current flows in the direction and is proportional to the magnetic field. (b) Demonstrates the proposal by Stephanov \textit{et al.} to observe the CVE in chiral matter triggered by a global rotation. In this case, the angular momentum $\bsg_{\text{rot}}$ imitates the role of the external magnetic field, thus inducing a net charge current along with it. In (c), we depict the tCVE where the curl of the tilt velocity emerges as an effective magnetic field. By tuning the system's parameters, the vorticity can be chosen in the same direction as the external electric field, thus giving rise to the tCVE and a net current flow in the direction parallel to the electric field and the vorticity. }
\label{fig:cve}
\end{figure}

Inside an isolated Weyl cone and long intervalley scattering times, the terms on the right-hand side of the kinetic equation (\ref{conti}) represent quantum anomalies and are due to the anomalous velocity and the Berry fields $\mathfrak{b}={\hat{\bp}}/{2|\bp|^2}$ and clearly break the charge conservation. The first term in the continuity equation (\ref{conti}) points to the fact that, in an individual Weyl valley with the monopole charge $\chi$, the chiral charge in the presence of the external fields is not invariant. This indicates the well-known chiral anomaly pertinent to the chiral fermions with a 3D Dirac spectrum\cite{nielsen1983adler}. As it is evident from Eq. (\ref{conti}), the nonconservation of the chiral charge can be remedied by the intervalley scatterings, as a chiral pump back to the opposite valley that takes place in the direction of the magnetic field.\cite{son2013chiral,spivak2016magnetotransport} This charge flow in the direction of the magnetic field can measure the magnitude of this effect and is the hallmark of the chiral anomaly that leads to the chiral magnetic effect (CME): the electric conduction due to the applied magnetic field (Fig. \ref{fig:cve}).

The last term in Eq. (\ref{conti}) is the main finding of this paper that describes an effect in tilted Weyl semimetals that, to the best of our knowledge, has not been noticed in previous studies. This term, coined here \textit{vortical anomaly}, introduces an anomalous effect from the inhomogeneous external perturbations and contributes to the chiral anomaly. This anomaly term arises from the interplay between electromagnetic fields and a geometric component such as vorticity, which drives the chiral vortical effect. {The vortical anomaly has a non-trivial origin that is rather distinct from the conventional chiral anomaly which presupposes an isotropic phase-space\citep{dwivedi2013classical}. While the chiral anomaly is related to a singularity in momentum space, i. e., the Berry monopole, the vortical anomaly is generated by the nontrivial texture of the inhomogeneous media wherein Weyl fermions move. These inhomogeneous media can be characterized by a parameter space where the texture, e. g., the tilt profile, has a vortex-like structure\citep{PhysRevLett.106.161102} analogous to a pseudo magnetic field which must naturally be formed from a parametric source term.}

{Notice furthermore that a similar approach has been proposed in Refs. \onlinecite{PhysRevD.90.105004,PhysRevD.88.025040}, where a torsional anomaly has been discussed in the framework of teleparallel gravity. In this case, space-time is mimicked by a lattice model in which the effect of deformations of $u_{ij}$ on free Dirac fields has been investigated. However, the obtained torsional anomaly is different from our tilt-induced anomaly since the authors considered non-tilted (particle-hole symmetric) fields, and no tilt has been generated in this case by the deformations.}

{The vorticity} is the property of the particles' motion (in relativistic fluid) and a well-defined quantity in the hydrodynamic limit \cite{deng2016vorticity}. The vorticity-induced current also was discussed by Son \textit{et al.}\cite{PhysRevLett.103.191601} in the hydrodynamic limit. In Weyl semimetals, Stephanov \textit{et al.}\cite{PhysRevLett.109.162001} pointed out that the vorticity-induced current can readily be realized in the nonequilibrium limit when the dynamics are due to the weak external fields (Fig. \ref{fig:cve}). In this limit, the vorticity can be reinterpreted as the angular velocity of the rotation of the local particle with respect to the laboratory. Thus, by identifying the Coriolis and the Lorentz forces and making the substitution $(\upmu/v_F^2)\:\bsg\rightarrow e\:\mathbf{B}$, it is possible to generalize the concept of the chiral vortical effect, at zero temperature, for the Weyl semimetals in a rotating frame.

{In deriving the continuity equation, we implemented the source-free gauge condition $\nabla_\br\cdot\bxi=0$. {As we have already mentioned in Sec. \ref{sec. 2}, this condition coincides physically with that of volume-preserving strain fields. Indeed, the divergence of the strain field is proportional to the divergence of the displacement field, which is itself given by a relative volume change, $\nabla\cdot\boldsymbol{\zeta}\propto\nabla\cdot\textbf{u}\propto \Delta V/V$}. This is physically interpreted as the vanishing of the symmetric part of the strain tensor associated with the rate of the change in the sheer volume in the continuum models\cite{abrikosov2017fundamentals}.}

Contrary to this, however, Eq. (\ref{conti}) suggests that the vorticity can be an internal property of the chiral Weyl particles' motion required by letting the tilt parameter be local. Therefore the chiral vortical effect for Weyl semimetals yields an anomalous effect that stems from the local tilt parameter without external rotation (Fig. \ref{fig:cve}). 

\section{Vortical response}\label{sec. 5}
\subsection{Chiral vortical effect}

In this section, we consider the electronic response of a tilted Weyl semimetal and study two transport phenomena that the vorticity can induce, namely, the chiral vortical effect and the anomalous vortical Hall effect. We first solve the equation set (\ref{anom}) for the velocity, which returns.
{
\begin{align}\label{eq:vel1}
\sqrt{G}\:\dot{\br}&=\mathbf{v}_\bp+\eta\:\mathbf{v}_t-\chi\:e\: \mathbf{E}\times\mathfrak{b}+\chi\:e\:(\mathfrak{b}\cdot\mathbf{v}_\bp) \mathbf{B}\nonumber\\
&\qquad+\chi\eta(\bp\times\bsg)\times\mathfrak{b}.
\end{align}}
It is evident from this equation that, except for the first two terms, which are simply the overall band velocity, all other terms are corrections due to the Berry curvature. Moreover, we note that the part of the anomalous velocity, which is just due to the vorticity, reads as $\mathbf{v}_{\bsg}=[\bsg-(\bsg\cdot\hat{\bp})\hat{\bp})]/2p$ and induces a dipole-moment structure in momentum space.
Next, substituting Eq. (\ref{eq:vel1}) into the current formula (\ref{curr}) for the isotropic momentum distribution at equilibrium, we find
{
\begin{align}
\mathbf{J}=\frac{\chi\:e^2 \mathbf{B}}{4\pi^2}\int_0^{\upmu_\chi} d\varepsilon\:f_{\text{eq}}+\frac{\chi\eta\:e\:\bsg}{6\pi^2}\int_0^{\upmu_\chi} \varepsilon d\varepsilon\:f_{\text{eq}}.
\end{align}}
In the low-temperature limit, where the Fermi-Dirac distribution is reduced to the step function $f_{\text{eq}}=\theta(\upmu - \varepsilon)$, the non-dissipative currents in response to the magnetic field and the vorticity are
\begin{align}\label{CME}
\mathbf{J}_{_\text{CME}}&=\chi\frac{e^2\: \upmu_\chi}{4\pi^2}\:\mathbf{B},\\
\label{CVE}
\mathbf{J}_{_{\bsg}}&=\chi\eta\frac{e \:\upmu_\chi^2}{12\pi^2}\:\bsg.
\end{align}

Both of these equations are non-dissipative equilibrium currents and robust against disorder as their coefficients have no dependency on the relaxation time.
Equation (\ref{CME}) is the standard chiral magnetic effect as the response of the Weyl fermions to the applied external magnetic effect\citep{son2013chiral,PhysRevLett.109.162001}. On the other hand, Eq. (\ref{CVE}) describes the tCVE proportional to the vorticity $\bsg$ that stems from an effective lattice modification via an externally induced strain. Here, we shall distinguish this result from the vortical effect caused by the global rotation\cite{dayi2017semiclassical} or by circularly polarized photons.\cite{PhysRevD.96.051902} In the latter two situations, the angular velocity of the frame, which may be viewed as an effective magnetic field,\cite{PhysRevLett.109.162001} causes the chiral current to flow along the axis of the rotation. In contrast, the main deriving force of the chiral current in Eq. (\ref{CVE}) is the vorticity tensor $\omega_\ell=\varepsilon_{i j\ell} \partial_i\zeta_j$, caused by the inhomogeneous tilt and the lattice deformations.

\subsection{Anomalous vortical Hall effect}

Let us now solve the Boltzmann equation (\ref{boltz}) by linearizing the distribution function in terms of the electric field (see Appendix). Particularly, for a stationary and homogenous system, the non-equilibrium part gives
\begin{align}\label{AVHE}
\delta f=\frac{e\tau\left[\mathbf{E}+ \tau\left(\omega_c\frac{\mathbf{B}}{|\mathbf{B}|}+\eta\:\bsg\right)\times\mathbf{E}\right]\cdot \mathbf{v}\left(\frac{\partial f}{\partial\varepsilon}\right)}{1+\tau^2\omega_c^2},
\end{align}
where $\omega_c=ev^2B/\varepsilon$ is the cyclotron frequency. One immediately notices the resemblance between the cyclotron frequency $\omega_c$ and the vorticity $\bsg$ in the second and third terms, respectively. This implies the emergence of a pseudomagnetic field proportional to $\mathbf{B}_5=\eta({\varepsilon_F}/{ev_F^2})\:\bsg$, thus suggesting an unusual transverse effect in the absence of the external magnetic field $\mathbf{B}$. The response is in agreement with the gravitomagnetic interpretation\citep{ryder1996quantum} of the tilt-induced force, similar to the response triggered when quasiparticles move in a (pseudogravitationally) inhomogeneous medium.

If we now incorporate this non-equilibrium distribution into the current formula (\ref{curr}) and consider the weak magnetic regime $\omega_c\tau\ll 1$, the anomalous transverse current and conductivity induced by the vorticity $\bsg$ can be computed as 
\begin{align}\label{cur_vort}
\mathbf{J}_{\bsg} &=\frac{\eta\:e\tau^2\:n}{\hbar}\;\frac{\upmu_\chi}{k_F^2}\:\bsg\times\mathbf{E},\\
\label{cur_vort2}
\frac{\upsigma_{ij}}{\upsigma_0}&=\eta\:\varepsilon_{ij\ell}\:\omega_\ell\tau,
\end{align}
where $n=\int d\bp/(2\pi)^3$ is the charge density. Furthermore, we identify $\upmu\approx \varepsilon_F$ in the low-temperature regime, and $\upsigma_0=\frac{e^2\tau n}{\hbar}\frac{\upmu}{k_F^2}$ is the Drude conductivity. This effect involves the mixing between the vorticity and the electric field and indicates the contribution to the anomalous Hall effect due to the tilt, \textit{i.e.}, $\upsigma_{ij}\propto(\partial_i\zeta_j-\partial_j\zeta_i)\tau$. A similar effect due to the response of Weyl semimetals to lattice dislocations has also been discussed in other studies\cite{huang2019torsional,PhysRevLett.116.166601}. Unlike the chiral magnetic and vortical effects, which are equilibrium currents, the anomalous vortical Hall effect in Eq. (\ref{AVHE}) is a non-equilibrium property and sensitive to the nature of the impurities, and, therefore, relatively smaller than the intrinsic anomalous Hall effect.

Equation (\ref{cur_vort2}) suggests that in the presence of pairs of TR symmetric nodes, the vortical Hall effect vanishes simply due to the cancellation of the opposite contributions from the symmetric cones, once one takes the sum over $\eta=\pm$. Therefore, breaking time-reversal symmetry and selective pumping of a specific Weyl node is critical for observing the transverse vortical effect. We elucidate this point in the following.

\subsection{Optical polarization driven net vortical Hall effect}

\begin{figure}[t]
	\centering
	\includegraphics[width=.9\linewidth]{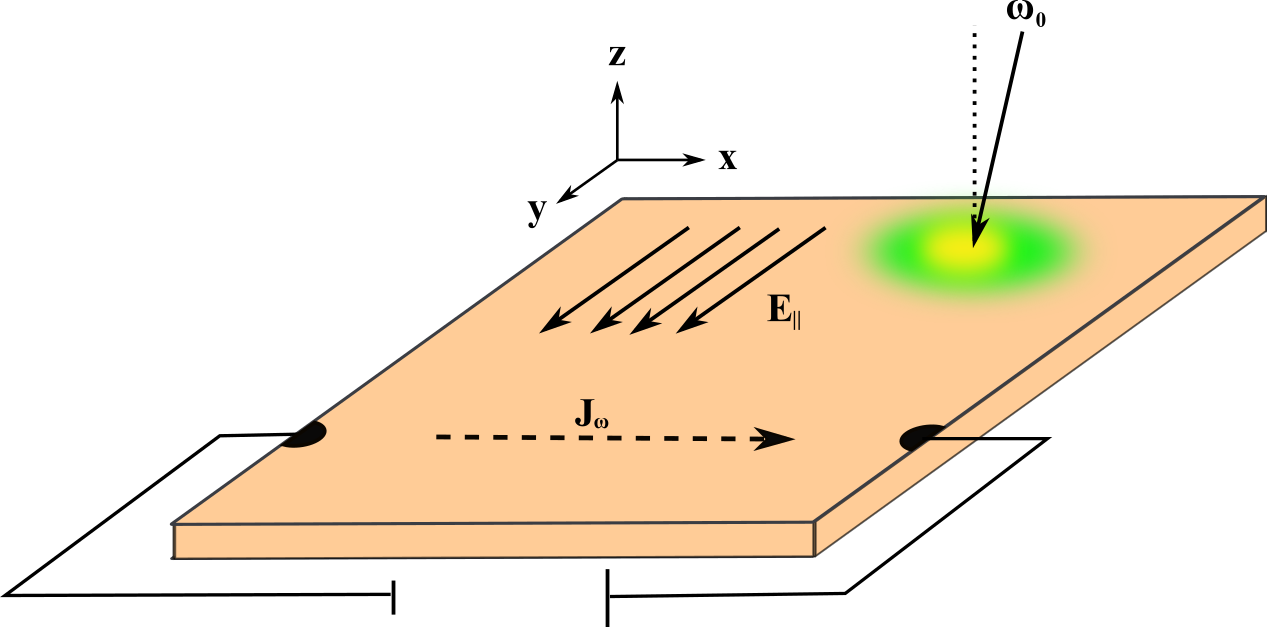}
	\caption{(Color online) {Tilt-induced vortical Hall effect can be realized in Weyl semimetal perturbed by the circularly polarized optical fields. An incident light pulse with polarization in the $\hat{\mathbf{y}}$ direction and frequency $\omega_0$ arrives, almost, normal to the sample. Due to the coupling of the electric field to the vorticity pseudovector $\bsg$, a transverse current $\mathbf{J}_{\bsg}$ is generated.}}
	\label{fig:cve_Hall}
\end{figure}

Circularly polarized light pulses provide suitable tools for controlling the electronic dynamics in topological materials. This is mainly due to the helicity of the polarized light pulses that, as an internal degree of freedom, can interact with and couple to the degrees of freedom of the Dirac fermion and thus unveil its topological nature. This is the primary mechanism behind the application of the polarized pulses in the selective pumping of a distinct valley in multi-valley topological materials. As such, the orbital angular momentum carried by the circularly polarized photons can couple to the vorticity, as it is given in Eq. (\ref{cur_vort}), producing a net dynamical vortical effect (see Fig. \ref{fig:cve_Hall} ). 

In order to analyze the light-matter interaction, we consider the general form of the Hamiltonian in the 4-vector notation of the Sec. \ref{sec. 2} as $H=\sum_{a=0}^3\mathbf{v}_a\sigma^a\cdot\bp$, and then apply a strong external magnetic field (along the $z$-direction) to the system where the spectrum can be constructed by minimal substitution $\bp\rightarrow\bp+e \mathbf{A}$.  To analyze this further within the semiclassical picture, we decompose the total vector potential into the static and oscillatory parts $\mathbf{A}=\mathbf{A}_\text{dc}+\mathbf{A}_\text{ac}$, where the $ac$ part is given by a circularly polarized electromagnetic radiation generated by the time-dependent potential with components $\mathbf{A}_\text{ac}(t)=\frac{E_0}{\omega_0}\: e^{i\omega_0t}(1,e^{i\phi},0)$. Here $\omega_0$ is the photon energy, $E_0$ is the field intensity, and $\phi$ gives the angle of polarization with $\phi=\pm\pi/2$, with its sign indicating right and left polarization, respectively. 
Consequently, the light-matter Hamiltonian gives
\begin{align}
 H_\text{ac}&=ev_F(\bar{\sigma}\cdot\mathbf{A}_\text{ac}+\eta\:\sigma_0\bxi\cdot\mathbf{A}_\text{ac})\nonumber\\
 &=H^\phi+\delta H_\text{tilt}
 \end{align}
 where $\bar{\sigma}=(\eta\:\sigma_x,\sigma_y,\chi\eta\:\sigma_z)$, and the free and the tilted Hamiltonians are given respectively as
 \begin{align}
H^\phi&=ev_F\frac{E_0}{\omega_0}\sigma_\phi e^{-i\omega_0t},\\
\delta H_\text{tilt}&=ev_F\frac{E_0}{\omega_0}\eta\zeta_\phi\sigma_0e^{-i\omega_0t},
\end{align}
where $\sigma_\phi=\eta\:\sigma_x+e^{i\phi}\sigma_y$ and $\zeta_\phi=\zeta_x+e^{i\phi}\zeta_y$.  One notices from $T H_\eta^\phi(t)T^{-1}=H_{-\eta}^{-\phi}(-t)$ that the Hamiltonian is odd under time-reversal symmetric transformation $T$. However, the overall Hamiltonian becomes periodic $H(t)=H(t+\text{T})$ with the periodicity $\text{T}=2\pi/\omega_0$, in which the system has discrete time-reversal symmetry. Moreover, the ground state of the tilt-free Hamiltonian $H^\phi$ is completely polarized for the right $\phi=+\pi/2$ and the left $\phi=-\pi/2$ polarizations. This is evident from the form of the Hamiltonian where for the right and the left polarization we have $H^{\circlearrowright/\circlearrowleft}\propto(\sigma_x\pm i\sigma_y)$, respectively, indicating that by applying a circular light polarization one can populate selectively a Weyl node with a particular chirality. The ground state of such a system describes the Weyl fermions with a distinct chirality, therefore pumping a specific valley\citep{tchoumakov2016magnetic,sari2015magneto,staalhammar2020magneto}.
The tilt effect can be implemented through $\delta H_\text{tilt}$, and since the amplitude of the matrix elements, on average, scales as $|\bxi|^2\ll 1$ therefore, the internode transitions due to the tilt can be ignored entirely.

By pumping a single Weyl node through the circular polarization, the effect of the tilt on the dynamics around that node can be explored by applying the semiclassical Boltzmann theory for a system perturbed by the time-dependent low-bias fields. To evaluate the transport coefficient in the presence of the external time-dependent electric field, we take into account the time-dependent modulation of the Boltzmann distribution (fermionic distribution) and express it in terms of the Fourier components as\cite{PhysRevB.94.245121}
\begin{align}
f=f_0+f_1\;e^{i\omega_0t}+f_2\:e^{2i\omega_0t}+\cdots.
\end{align}

Next, solving the Boltzmann equation recursively for different Fourier components (see Appendix B) the first harmonic gives the non-equilibrium solution as $\delta f=f_1(\omega_0;\bsg,\mathbf{E},\mathbf{B})\:e^{i\omega_0t}\left(\frac{\partial f}{\partial\varepsilon}\right)$ such that\cite{PhysRevB.95.085127}
\begin{align}\label{sol_1}
 f_1&= \frac{\tau e
\Big[(1-i\tau\omega_0)\:\mathbf{E}+\tau\:\mathbf{E}\times(\eta\:\bsg+\omega_c\:\hat{\mathbf{B}})\Big]\cdot\mathbf{v}}{(1-i\omega_0\tau)^2+\tau^2(\eta\:\bsg+\omega_c\:\hat{\mathbf{B}})^2}.
\end{align}
\begin{figure}[ht!]
	\centering
	\includegraphics[width=.9\linewidth]{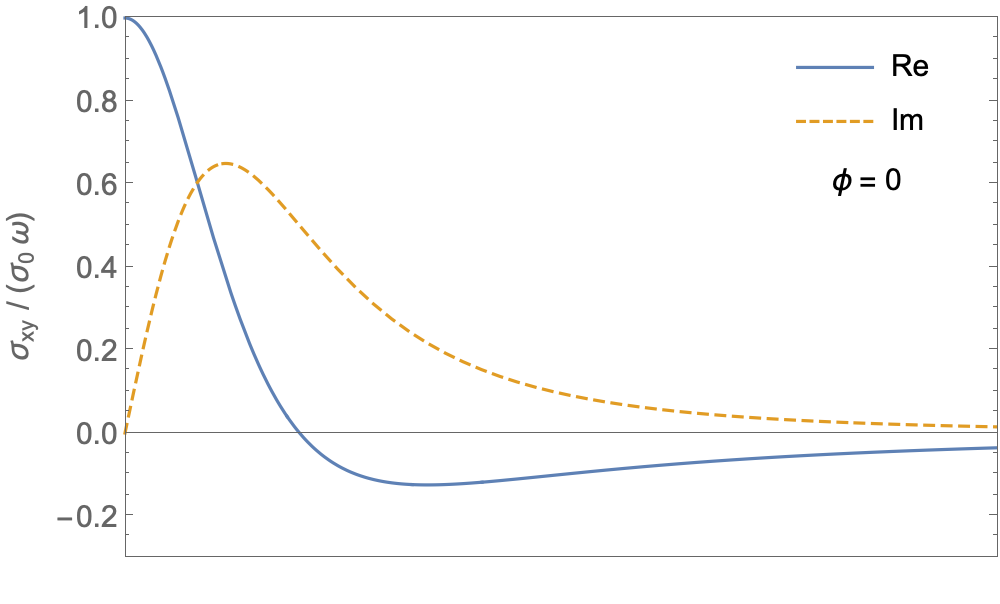}
	
		\vspace{-11pt}	
		
	\includegraphics[width=.91\linewidth]{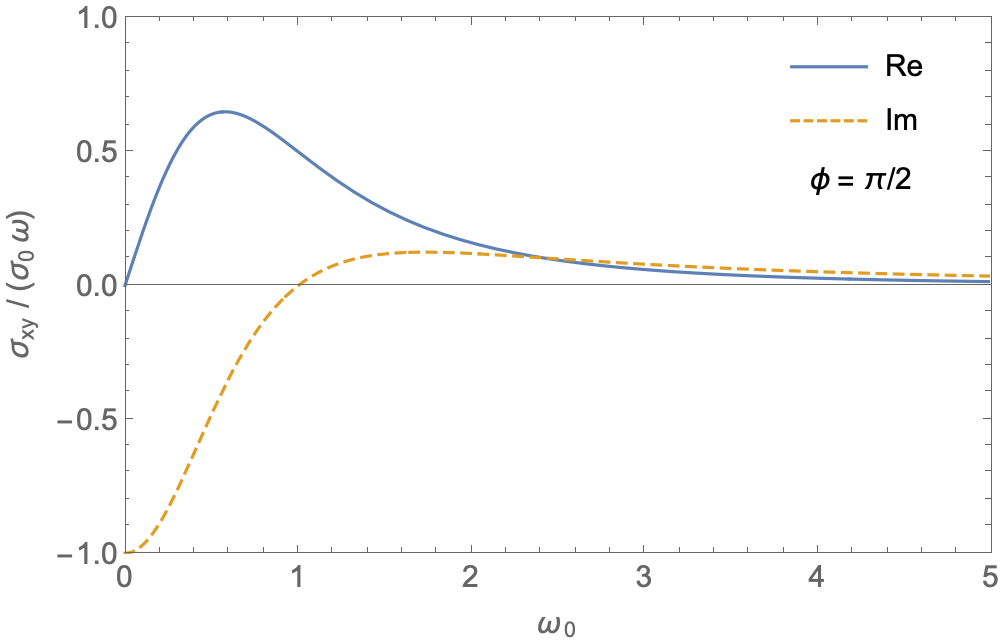}
	\caption{(Color online) (Top panel) The transverse vortical conductivity due to the tilt as a function of the frequency $\omega_0$. For the plane or linear polarization (top panel) with angle $\phi=0, \pi$, the real part of the total conductivity shows Drude behavior starting from a maximum and vanishes at around the resonant frequency. (Bottom panel) the same transverse conductivity as a function of the $\omega_0\tau$ for the circular polarization $\phi=\pi/2$ showing a Lorentzian behavior and reaches a maximum as the resonant frequency.}
	\label{fig:Hall_cond}
\end{figure}
By rescaling and rotating the momenta and transforming the Hamiltonian, it is always possible to project the tilt velocity onto a certain direction\cite{tchoumakov2016magnetic,rostamzadeh2019large}. {Let us use a tilt profile that is reminiscent of the Landau gauge with the position-dependent components as $\zeta_x(y)$ and $\zeta_y(x)$. Previous studies\citep{PhysRevLett.122.056601,yekta2021tune} assumed inhomogeneous tilt parameters with step-like profile as $\zeta_i(x_j)\propto\theta(x_j)$ where $\theta$ is the step function. Here, however, we opt for a smooth profile by using instead a hyperbolic function, $\zeta_i(x_j)=\omega\:\tanh(x_i)$, that can be linearized as $\zeta_i(x_j)\approx \omega\:x_j+O(x_j^2)$ to give constant vorticity vector in a preferred direction as $\bsg_\eta=\eta\omega\:\hat{\mathbf{z}}$ and $\omega$ is the modulus of the vorticity vector.} Using the solution (\ref{sol_1}) the total tilt-induced complex Hall conductivity then gives
\begin{align}\label{complex_cond1}
\frac{\text{Re}\: \upsigma_{xy}}{\upsigma_0} &= \frac{(1-\omega_0^2\tau^2)\cos\phi+2\omega_0\tau \sin\phi}{(1+\omega_0^2\tau^2)^2}\;\omega,\\
\label{complex_cond2}
\frac{\text{Im}\: \upsigma_{xy}}{\upsigma_0}  &= \frac{2\omega_0\tau \cos\phi-(1-\omega_0^2\tau^2)\sin\phi}{(1+\omega_0^2\tau^2)^2}\;\omega,
\end{align}
in the limit of low magnetic fields $\omega_c\ll\omega_0$. First note that in the static limit $\omega_0\approx 0$ and $|\bsg|\ll\omega_c$, we identically restore the semiclassical result (\ref{cur_vort}) and (\ref{cur_vort2}), where the real and imaginary parts of $\upsigma_{xy}/\upsigma_0\omega$ are simply given by $\cos\phi$ and $\sin\phi$, respectively.
The net transverse conductivity is thus maximal for the circular polarization when $\phi=\pm\pi/2$, while the amplitude is zero for the linear and planar polarization $\phi=0,\pi$. 
More interesting is the behavior of the complex conductivity under light polarization. Considering a linear polarization where $\phi=0$, the real part of the Hall conductivity, $\text{Re} \:\upsigma_{xy}$, shows Drude-type behavior, and the peak decreases monotonically by increasing the frequency, which is also observed in the strain modulated WSMs\cite{heidari2020chiral}. The imaginary part of the conductivity, which is proportional to the dielectric constant, also shows behavior in full agreement with the Drude model\cite{ashcroft1976solid} (Fig. \ref{fig:Hall_cond} upper panel). 

For the circular polarization, however, the conductivity obeys a Lorentz-like pattern rather than a Drude behavior. This means that for $\omega_0\tau\ll 1$ the real part of the conductivity exhibits a linear behavior and vanishes at $\omega_0=0$ while it shows a Lorentz peak at the resonance point $\omega_0\tau\approx 1$ (Fig. \ref{fig:Hall_cond} bottom panel). 
\begin{figure}
	\centering
	\includegraphics[width=.91\linewidth]{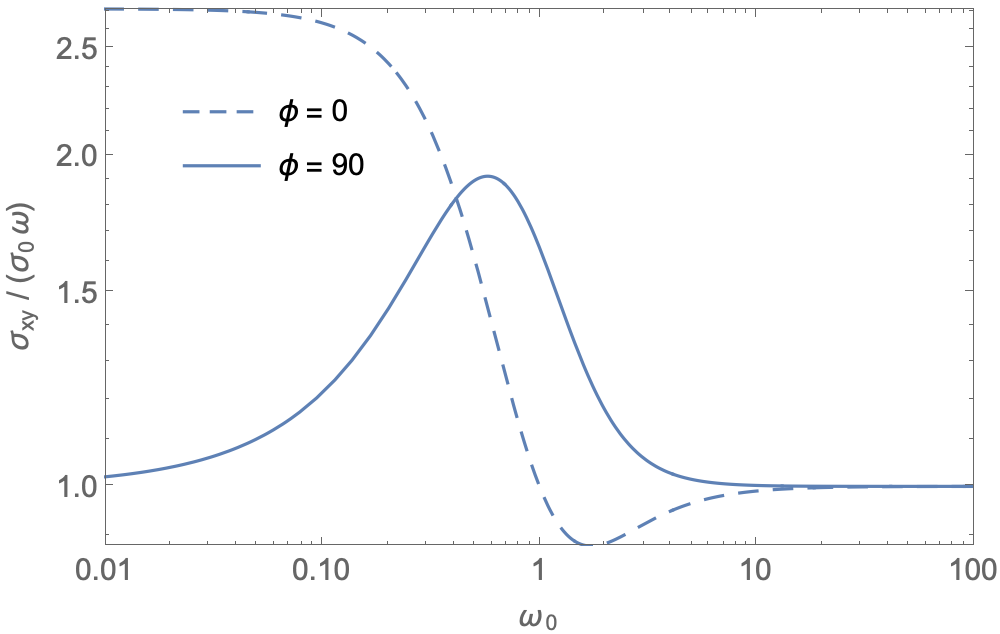}
	\caption{(Color online) The non-Drude behavior of the real part of the optical conductivity for circular polarization in logarithmic scale.}
	\label{fig:non_drude}
\end{figure}

This non-Drude behavior (see Fig. \ref{fig:non_drude}) is evident from the form of the equations for the complex conductivity. One assumes in general the conductivity to behave according to the Drude approximation using Eq. (\ref{sol_1}), i. e.,
\begin{equation}
\upsigma_{\text{Drude}}=\frac{1}{3}e^2\:v_F^3\:\int_0^{\varepsilon_F} d\varepsilon\:\rho(\varepsilon)\:\frac{\tau}{1-i\omega_0\tau} \left(\frac{\partial f_0}{\partial\varepsilon}\right),
\end{equation}
 where $\rho(\varepsilon)$ is the density of states. However, in a tilted system, in addition to the Lorentz force, one must consider the effect of the pseudomagnetic field [see Appendix B, Eq. (\ref{apx:Lorentz_force})], stemming from the spatial variation of the tilt, in the dynamics of the Weyl fermions. Next, by solving the Boltzmann equation perturbatively in terms of the electric field we capture the conductivity that is proportional to $e^{i\phi}/(1-i\omega_0\tau)^2$, as it is clear from Eqs. (\ref{complex_cond1}) and (\ref{complex_cond2}), such that in the low-frequency limit, one has $\lim_{\omega_0\rightarrow 0} \text{Re}\upsigma(\omega_0)\approx \omega_0\tau$.

\section{Conclusion}
\label{sec. 6}
In summary, we have studied the effect of a spatially varying tilt in Weyl semimetals. In the case of a non-zero rotational of the tilt field, $\nabla\times \mathbf{v}_t(\br)\neq 0$, we find a quantum anomaly different from the usual chiral one. We have shown that for a specific profile of the tilt parameter with nonzero curl, the particles around the anisotropic cone experience a Coriolis force proportional to the vorticity of the tilt field. Hence the vortical anomaly, as the coupling between the electromagnetic field and the vorticity, is to great extent reminiscent of the chiral anomaly where the vorticity of the tilt field, $\bsg=\nabla\times\bxi(\br)$, plays the role of an effective magnetic field.
This anomaly naturally gives rise to global transport effects, just as the chiral anomaly gives rise to the chiral magnetic effect.  

Using the semiclassical equations of motion, we demonstrated that the vorticity indeed gives rise to the chiral vortical effect: a current flows in the direction of the vorticity field, analogous to the chiral magnetic effect. Although the proposed vortical effect shares similarities with the chiral vortical effect in the hydrodynamic limit, it is distinct from the rotation-induced axial charge flow in Weyl semimetals\cite{PhysRevLett.109.162001}. The chiral vortical effect is well defined in the hydrodynamic limit for the relativistic fluid having vortex currents where they exert a global rotation on segments of the systems. We furthermore disclose a transverse effect due to the tilt, which arises from the vorticity of the tilt field and yields a current in a direction perpendicular to the electric field, just as the anomalous Hall effect. These two transport phenomena, namely the chiral vortical effect and vortical Hall effect, are the main results of this paper, and they are specific to Weyl semimetals with spatially varying tilt. 

As there are an even number of contact points in the Weyl semimetal, the experimental observation of this form of Hall effect due to the tilt would require a mechanism to avoid the cancellation of the effect by opposite currents arising from the opposite nodes. One, therefore, needs to break TR symmetry and occupy preferentially Weyl nodes with the same chirality. We demonstrated that using the optical field with circular polarization the transverse vortical conductivity reaches a maximum at a frequency comparable with the effective cyclotron frequency of the vortical field. Furthermore, this condition indicates the cyclotron resonance in the pseudo-Landau levels due to the vorticity as the effective magnetic field.

\textit{Note added.} Recently, we came across two related works\citep{haller2022black,konye2022anisotropic} that compute similar transverse deflection in the trajectories of the Weyl particles affected by the position-dependent tilt parameter.

\begin{acknowledgements}
This study was funded by Scientific Research Projects Coordination Unit of Istanbul University, Project NO M2019-34733. We furthermore acknowledge financial support from Agence Nationale de la Recherche via the ANR project DIRAC3D (Grant No. ANR-17-CE30-0023).
\end{acknowledgements}

\appendix
\section{Time-reversal invariant WSM}

Let us first start with the Hamiltonian for two Weyl nodes with opposite chirality, in the absence of time-reversal  symmetry
\begin{align}\label{eq:WH2}
H=v\:\bp_{||}\cdot\sigma+\left(\Delta-\frac{k_z^2}{2m}\right)\sigma_z,
\end{align}
giving two Weyl nodes at $(0, 0, \pm p_z^0)$ with $p_z^0=\sqrt{2m\Delta})$. The momentum in the $z$-direction is now defined with respect to those points, $k_z=p_z^0+p_z$ and $\bp_{||}=(p_x,p_y)$. Expanding the Hamiltonian around the Weyl nodes gives the low energy approximation as
\begin{align}
H_{\text{low}}=v\:\bp_{||}\cdot\sigma+\chi \:v_zp_z\sigma_z,
\end{align}
with the velocity $v_z=\sqrt{{2\Delta}/{m}}$ and $\chi=\pm$ is the chirality index. Note that the Berry curvature for this system comes with two chiralities
\begin{align}
\mathfrak{b}_z^\chi=\frac{\chi\:p_z v^2v_z}{2(v^2 p_{||}^2+v_z^2p_z^2)^{3/2}}.
\end{align}
It is clear that the Hamiltonian (\ref{eq:WH2}) does not respect time-reversal symmetry, through the condition $T H(\bp)T^{-1}=H^\ast(-\bp)$, and to do so, there must be a second pair of Weyl nodes with opposite chirality, e.g., in the $xz$-plane. To keep track of this second pair of Weyl nodes, we introduce another index $\eta$, so that the final low-energy Hamiltonian reads as
\begin{align}
H_{\chi,\eta}=v(\eta \:p_x \sigma_x+p_y\sigma_y)+\eta \:\chi \:v_zp_z\sigma_z,
\end{align}
giving rise to four contact points $(\eta\; p_x^0, 0,\chi\eta\; p_z^0)$ identified via the chirality index $\chi=\pm 1$ and the TR-index $\eta=\pm$. The Berry curvature for this system gives
\begin{align}
\mathfrak{b}_z^\chi=\frac{\eta^2\:\chi\:p_z v^2v_z}{2(v^2 p_{||}^2+v_z^2p_z^2)^{3/2}}
\end{align}
and is thus independent of the index $\eta$. Therefore, with our choice $p_y=0$ for the zero-energy plane, the Weyl points $\eta(p_x^0,0,p_z^0)$ carry topological charge $\chi=+1$ whereas the nodes $\eta(p_x^0,0,-p_z^0)$ are associated with the charge $\chi=-1$. We then perturb the Weyl cone and induce tilt, where it has opposite signs in the two time-reversal symmetric nodes given by
\begin{align}
 H_\text{tilt}=\eta\:\mathbf{v}_t\cdot\mathbf{p}.
\end{align}

\section{ Circular polarization and the first harmonics}

We suppose that the electric field is time-dependent and given by $\mathbf{E(t)}=\mathbf{E}_\chi\:e^{i\omega_0t}$. The solution of the Boltzmann equation can be perturbatively expressed in terms of the harmonics as
\begin{align}
f=f_0+f_1\;e^{i\omega_0t}+f_2\:e^{2i\omega_0t}+\cdots,
\end{align}
but in this work, our focus is only up to the first harmonic. The force term in Eq. (\ref{fors}) reads as
\begin{align}\label{apx:Lorentz_force}
\sqrt{G}\dot{\mathbf{p}}&=(-e\:\mathbf{E}+\eta[\bp\cdot\nabla]\zeta)-e\:\mathbf{v}\times\mathbf{B}-e^2\eta\chi\:(\mathbf{E}\cdot\mathbf{B})\mathfrak{b}\nonumber\\
&\hspace{1in}+\eta\:\bp\times\bsg.
\end{align}
Next, the Boltzmann equation expanded in linear terms in the electric field gives
\begin{align}
&(1/\tau-i\omega_0)f_1+\sqrt{G}\Big(-e\:\mathbf{E}-e^2\:\eta\chi(\mathbf{E}\cdot\mathbf{B})\mathfrak{b}\Big)\cdot\mathbf{v} \left(\frac{\partial f_0}{\partial\varepsilon}\right)\nonumber\\
&\hspace{.2in}+\sqrt{G}\Big(\eta[\bp\cdot\nabla]\zeta-e\:\mathbf{v}\times\mathbf{B}+\eta\:\bp\times\bsg\Big)\cdot\nabla_\bp f_1=0.
\end{align}
We furthermore assume that the nonequilibrium part can be expressed as $f_1=\mathcal{O}\cdot\mathbf{v} \left(\frac{\partial f_0}{\partial\varepsilon}\right)$ so that $\nabla_\bp f_1=\frac{v^2}{\varepsilon}\;\mathcal{O}\:\left(\frac{\partial f_0}{\partial\varepsilon}\right)$. This consequently simplifies the Boltzmann equation,
\begin{align}
&(1/\tau-i\omega_0)\mathcal{O}-\frac{v^2}{\varepsilon}(e\:\mathbf{B}+\frac{\eta\:\varepsilon}{v^2}\:\bsg)\times\mathcal{O}=\nonumber\\
&\hspace{1.5in}-e\:\mathbf{E}_\chi-\eta\:e^2\:(\mathbf{E}_\chi\cdot\mathbf{B})\mathfrak{b},
\end{align}
which is solved up to the first order in the electric field,
\begin{align}
\mathcal{O}=\frac{\tau e
\Big[(1-i\tau\omega_0)\:\mathbf{E}_\chi+\tau\:\mathbf{E}_\chi\times(\eta\:\bsg+\omega_c\:\hat{\mathbf{B}})\Big]}{(1-i\omega_0\tau)^2+\tau^2(\eta\:\bsg+\omega_c\:\hat{\mathbf{B}})^2}.
\end{align}

\bibliographystyle{apsrev4-1}
\bibliography{referal} 

\end{document}